
\input phyzzx
\hoffset=0.2truein
\hsize=6truein
\voffset=0.1truein
\def\TITLEPAGE{\frontpagetrue}
\def\IASSNS#1{\hbox to\hsize{\tenpoint \baselineskip=12pt
        \hfil\vtop{
        \hbox{\strut iassns-hep-94-#1}
}}}
\def\PUPT#1{\hbox to\hsize{\tenpoint \baselineskip=12pt
        \hfil\vtop{
        \hbox{\strut PUPT-94-#1}
}}}
\def\HEP{\hbox to\hsize{\tenpoint \baselineskip=12pt
        \hfil\vtop{
        \hbox{\strut hep-ph/9501396}
}}}
\def\SUNYSB{
\centerline{$^*$
 Institute for Theoretical Physics, State University of New York}
\centerline{Stony Brook, New York 11794-3840}}
\def\PRINCETON{
\centerline{$^\dagger$Physics Department, Princeton University}
\centerline{Princeton, New Jersey 08544}}

\def\TITLE#1{\vskip .0in \centerline{\fourteenpoint #1}}
\def\AUTHOR#1{\vskip .1in \centerline{#1}}

\def\ABSTRACT#1{\vskip .1in \vfil \centerline{\twelvepoint
\bf Abstract}
   #1 \vfil}
\def\ENDTITLEPAGE{\vfil\eject\pageno=1}
\hfuzz=5pt
\tolerance=10000
\TITLEPAGE
\PUPT{1527}
\HEP
\TITLE{$\Omega _0<1$ From Inflation}

\vskip 20pt
\vskip 20pt

\AUTHOR{Martin Bucher$^{\dagger }$\footnote\ddagger{Talk
presented at the ASTPART conference in Stockholm, Sweden
in September 1994.}}
\AUTHOR{Alfred S. Goldhaber$^*$}
\AUTHOR{Neil Turok$^\dagger$}
\vskip 20pt
\PRINCETON
\vskip 20pt
\SUNYSB
\nobreak
\ABSTRACT{
An inflationary scenario that leads to $\Omega _0<1$
today is presented. An epoch of `old' inflation during which
the smoothness and horizon problems are solved is followed
by a shortened epoch of `new' inflation. Old inflation
exits through the nucleation of a single bubble, leading
to negative spatial curvature on slices of constant cosmic
time. We calculate the spectrum of density perturbations in
such a scenario.
}
\rightline{January 1995}

\ENDTITLEPAGE

\eject

\def\sech{ {~\rm sech ~ }}
\def\H{{\cal H}}

\REF\bgt{M. Bucher, A.S. Goldhaber and N. Turok,
``An Open Universe From Inflation,"
Princeton Preprint (10-94) hep-ph 94-11206}
\REF\deluccia{
S. Coleman and F. De Luccia, ``Gravitational Effects on and of
Vacuum Decay," Phys. Rev. {\bf D21,} 3305 (1980).}
\REF\gott{
J.R. Gott, III, ``Creation of Open Universes from de Sitter Space,"
Nature {\bf 295,} 304 (1982); J.R. Gott, III, and T.S. Statler,
Phys. Lett. {\bf 136B,} 157 (1984);
J.R. Gott, III, ``Conditions for the Formation of Bubble Universes,"
in E.W. Kolb et al., Eds. {\it Inner Space/Outer Space,} (Chicago,
U. of Chicago Press, 1986).}
\REF\lyth{
D. Lyth and E. Stewart, ``Inflationary Density Perturbations
with $\Omega <1,$'' Phys. Lett. {\bf B252,} 336 (1990).}
\REF\ratra{
B. Ratra and P.J.E. Peebles, ``Inflation in an Open Universe,"
PUPT-1444 (Feb. 1994);
B. Ratra and P.J.E. Peebles, ``CDM Cosmogony in an Open Universe,"
PUPT-1445 (Feb. 1994).}
\REF\allen{
B. Allen and R. Caldwell, Preprint in preparation.}
\REF\sasaki{
K. Yamamoto, T. Tanaka, and M. Sasaki, Preprint in preparation.}
\REF\bt{
M. Bucher and N. Turok, ``Power Spectrum from Open Inflation
for Arbitrary $m_{fv}^2/H^2_{fv},$" In preparation.}

\chapter{INTRODUCTION}

The ratio of the mean density of the universe to
the critical density $\Omega _0=\rho _0/\rho _{crit}$
is a crucial cosmological parameter that determines
the geometry of the universe. For $\Omega =1,$ slices
of constant cosmic time are Euclidean, with vanishing
spatial curvature. On the other hand, if $\Omega <1,$
slices of constant cosmic time are hyperbolic, with
constant negative spatial curvature. Although there
is some observational evidence
in favor of $\Omega =1,$ many observations
suggest a universe of subcritical density, with
$\Omega <1.$

Inflation is an attractive scenario for the very early
universe because of its elegant solution of the horizon
and smoothness problems. The exponential expansion during
inflation smooths out whatever inhomogeneities may have
existed prior to inflation. Moreover, in addition to
erasing initial conditions, inflation also provides
a predictive mechanism for generating the primordial
density perturbations that seed later structure formation.

In this contribution we investigate the compatibility of
$\Omega _0<1$ with inflation. In particular, we investigate
the power spectrum of adiabatic density perturbations
that results from a low-$\Omega _0$ model of inflation.
This work is discussed in more detail in ref. \bgt ,
which also contains a more extensive set of references.

Although standard inflation predicts $\Omega _0$
equal to one to extremely high accuracy, inflation
with low $\Omega _0$ is also possible. The basic idea
is due to Coleman and de Luccia\refmark{\deluccia }
and to Gott\refmark{\gott }. When the
false vacuum decays through the nucleation of a single
bubble, inside the bubble the spatial hypersurfaces
on which the scalar field is constant,
indicated by the solid lines in Fig. 1,
are surfaces of constant negative spatial curvature.
In new inflation the surfaces on which the inflaton field
is constant are the natural constant time hypersurfaces.
The interior of the bubble contains an expanding open
FRW universe, described by the metric
$$
ds^2=-dt^2+a^2(t)\cdot [d\xi ^2+\sinh ^2[\xi ]
d\Omega _{(2)}^2] .
\eqn\aaa$$
We call this patch region I. Another coordinate patch
(which we call region II) with the line element
$$
ds^2=d\sigma ^2+b^2(\sigma )\cdot [-d\tau  ^2+\cosh ^2[\xi ]
d\Omega _{(2)}^2]
\eqn\rgt$$
(i.e., a coordinate system somewhat akin to `Rindler' coordinates)
covers the exterior of the bubble and part of the interior
as well. For de Sitter space $a(t)=H^{-1}\sinh [Ht]$ and
$b(\sigma )=H^{-1}\sin [H\sigma ].$\footnote\dagger{
Actually to cover maximally extended de Sitter space
requires three additional coordinate patches identical
in structure to region I, but this point shall not
concern us here.}

Consider a potential for the inflaton field
of the form sketched in Fig. 2.
The inflaton field starts in the false vacuum during
the initial epoch of old inflation, during which the
horizon and smoothness problems are solved. During this time
the spacetime geometry is that of pure de Sitter space,
with no preferred time direction. Then old inflation is
exited through the nucleation of a single bubble. The nucleation
event, a classically forbidden process, roughly corresponds
to the region below the dashed line in Fig. 1. Then the
bubble expands classically, at a velocity approaching the
velocity of light. The thin wall picture is
inadequate for our purposes here---the bubble wall has
no well defined edge. Instead of tunneling directly
to the true vacuum, the inflaton field tunnels onto
a slowing rolling potential, so that a shortened
epoch of new inflation occurs inside the bubble.
Formally, $\Omega =0$ on the null surface bounding
region I (at $t=0$). As $t$ increases, $\Omega $
flows toward one until reheating, and thereafter
flows away from one.

The value of $\Omega _0$ today is determined
by the shape of the inflaton field potential. The
length of new inflation is proportional to the
logarithm of the deviation of $\Omega $ from unity
at reheating. Consequently, no unnatural fine tuning
is required to obtain a low value of $\Omega _0$
because it is the logarithm of a small number rather
than the small number itself that must be adjusted,
and the usual naturalness argument against $\Omega _0
\ne 0$ is inapplicable. In fact, it seems that inflation
is necessary to overcome the naturalness argument.

\chapter{DENSITY PERTURBATIONS}

In the scenario for open inflation described above,
Gaussian adiabatic density perturbations are
generated by the same physical mechanism as in
flat inflation. Vacuum quantum fluctuations of the
inflaton field, initially well within the horizon,
turn into adiabatic density perturbations as they
cross the horizon. However, the computation of the
power spectrum from open inflation is more complicated
because of the effects of curvature
on large scales. On small co-moving scales (small
compared to the curvature scale) one has an approximately
scale invariant spectrum of density perturbations,
just as in flat inflation, because those modes
cross the horizon during the latter part of the
new inflationary epoch, when $\Omega $ was close
to one and when there was an approximate time
translation symmetry of the physical horizon volume.
On the other hand, larger scales (of order the curvature scale)
cross the horizon during the earlier part
of new inflation, when $\Omega $ was significantly
less than one and when the approximate time translation
symmetry of the physical horizon volume
during inflation is badly broken. Moreover, these
large scale modes are more sensitive to the quantum
state of the scalar field during the prior old inflationary
epoch.

The starting point for a calculation of the power
spectrum from open inflation is the quantum state
of the perturbations of the inflaton field about
the background solution during old inflation. These
fluctuations are well described by a free scalar field
of mass $m^2=V^{\prime \prime }[\phi _{fv}]$ in a de Sitter
background. To simplify the computation, we shall
take $m^2=2H^2.$ It is well known that after a sufficient
amount of old inflation, an arbitrary initial state will
approach what is known as the Bunch-Davies, or Euclidean
vacuum, described by the two-point Wightman function
$$
G^{(+)}={H^2({1\over 4}-\nu ^2)\over 16\pi }
{}~~~\times {}_2F_1\biggl( {3\over 2}-\nu ,
{3\over 2}+\nu ;2;{I(X,X')+1\over 2}\biggr)
\eqn\aab
$$
where $\nu ^2={9\over 4}-m^2/H^2$ and $I(X,X')=
-\bar w\bar w^\prime +\bar u\bar u^\prime
+\bar x\bar x^\prime +\bar y\bar y^\prime
+\bar z\bar z^\prime
-i\epsilon \cdot \varepsilon(X,X^\prime ).$

The quantum fluctuations of the scalar field coupled
to scalar gravity at the end of inflation determine
the spectrum of adiabatic density perturbations seen
today. These fluctuations are calculated by propagating
the `positive frequency' modes of the Bunch--Davies
vacuum into the bubble interior through the bubble wall.
As the modes propagate through the bubble wall the
effective mass squared of the inflaton field changes, from
$V^{\prime \prime }[\phi _{fv}]$ outside the bubble
wall, to a negative value inside the bubble wall (near
the maximum of the potential), and finally to nearly
zero inside the bubble, where the potential
is very flat.

In order to calculate the effects of this changing
effective mass, it is necessary to work
in coordinates that maximally exploit the
$SO(3,1)$ symmetry of the expanding bubble
solution. We start in region II, described by
the line element in eqn. \rgt , and express
the Bunch-Davies two point function as a sum
of region II modes. To calculate the power
spectrum it is sufficient to consider only
the $s$-wave. The $s$-wave mode functions
$$
f^{(\pm )}(u,\tau ;\zeta )=
{1\over 4\pi \sqrt{\vert \zeta \vert }}
{e^{+i\zeta u}\over \sech [u]}{e^{\mp i\vert \zeta \vert \tau }
\over \cosh [\tau]}
\eqn\aac$$
form a basis of normalized modes for the $s$-wave sector in
region II, where $\tanh [u]=\cos [H\sigma ].$ We use
units in which $H=1$ so that $b(\sigma )=\sin (\sigma ).$
Although the mode functions $f^{(+)}(u,\tau ;\zeta )$
formally appear to be `positive frequency' mode functions
(when viewed in region II hyperbolic coordinates), the
true `positive frequency' mode functions corresponding
to modes that annihilate the Bunch-Davies vacuum are
the linear combinations
$$
g^{(+)}_\zeta = { e^{\vert \zeta \vert \pi /2} f^{(+)}_\zeta -
e^{-\vert \zeta \vert \pi /2} f^{(-)}_\zeta
\over  \left( e^{+\vert \zeta \vert \pi } -
e^{- \vert \zeta \vert \pi }
\right) ^{1/2} }
\eqn\mdf$$
related by a Bogolubov transformation.

Expanding the $s$-wave component of the inflaton field as a
sum over region II mode functions, one has
$$
\hat \phi _{(s)}=\int _{-\infty }^{+\infty }d\zeta ~
\left[ g^{(+)}_\zeta \hat a_\zeta +g^{(+)}_\zeta \hat a_\zeta ^\dagger
\right]
\eqn\aad$$
where the operators $\hat a_\zeta $ annihilate the Bunch-Davies vacuum.

Assuming for simplicity that the size of the bubble is small
compared to the Hubble length during old inflation and that
the bubble radius and thickness is small compared to the
co-moving length scales of interest, we make the approximation
that the mass changes discontinuously from $m^2=2H^2$ in region
II to $m^2=0$ in region I. To calculate expectation values in
region I, one propagates the mode functions given in
eqn. \mdf ~ into region I. This involves matching
on the null surface separating regions I and II. In
region I we take into account the coupling of the
inflaton field perturbations to scalar gravity
using the gauge invariant formalism, originally developed
by Bardeen. Instead of $\hat \phi $ it is convenient to
use as the dynamical variable the gauge invariant gravitational
potential $\hat \Phi ,$ regarded as a quantum field.
The power spectrum is the hyperbolic Fourier transform of the
two point function of  $\hat \Phi $ at constant time.

The density perturbations are most elegantly described in
terms of the variable
$$\chi ={2\over 3}{\H ^{-1}\Phi ^\prime +
\Phi \over 1+w}+\Phi
\eqn\aae$$
which in flat models is conserved on superhorizon scales.
In open models $\chi $ is conserved on superhorizon scales
while $\Omega $ is close to one---that is during the latter
part of new inflation until the latter part of matter domination,
when the universe starts to become curvature dominated.

In terms of $\chi $ the power spectrum is
$$P_\chi (\zeta )\sim \left( {H^3\over
V_{,\phi }}\right) ^2 \cdot
{\coth [\pi \zeta ]\over \zeta (\zeta ^2+1)}
\eqn\psopen$$
where
$$\langle ~\chi (\zeta )~\chi ({\zeta '}) ~\rangle =
P_\chi (\zeta )\delta (\zeta -\zeta ')
\eqn\tpf$$
and
$$
\langle \chi (\xi )~\chi (0)\rangle =\int _0^\infty d\zeta ~
\zeta ~
{\sin [\zeta \xi ]\over \sinh [\xi ]}~
P_\chi (\zeta )
\eqn\aaf$$
With these conventions $P\sim \zeta ^{-3}$ corresponds
to scale invariance. This is seen by computing
$\langle \chi^2(0)\rangle $ using the small $\xi$ limit of
eqn. \psopen ~
and noting the logarithmic divergence
for large $\zeta $.

The spectrum of density perturbations from open inflation
has also been considered by Lyth and Stewart\refmark{\lyth}
and by Ratra and Peebles\refmark{\ratra}. Their treatments
differ from ours in the choice of initial conditions.
Instead of using the Bunch-Davies vacuum as an initial condition in
the epoch of old inflation, they use a conformal vacuum
for region I considered in isolation.
Consequently, the power spectrum given
in eqn. \psopen ~
differs from that given in refs. \lyth ~ and \ratra
by the factor of $\coth [\pi \zeta ],$ precisely as one
would expect from the Bogolubov coefficients in
eqn. \mdf .

In related work Allen and Caldwell\refmark{\allen }
and Yamamoto, Tanaka,
and Sasaki\refmark{\sasaki }
independently obtained the same Bogolubov
coefficients for the Bunch-Davies vacuum in terms of
hyperbolic modes.

We have also investigated the more general case
where $m^2/H^2\ne 2.$ It turns out that for reasonable
values of $\Omega ,$ large enough to be consistent with
lower bounds on the observed mass density of the universe,
varying $m^2/H^2$ only slightly changes the power spectrum
for observationally accessible wave numbers.
These results will be presented elsewhere.\refmark{\bt }

We acknowledge useful conversations with Bruce Allen,
Robert Caldwell, David Lyth, Jim Peebles, Bharat Ratra,
Misao Sasaki, and Frank Wilczek.
This work was partially supported by NSF contract
PHY90-21984 and by the David and Lucile Packard foundation.

\centerline{\bf Figure Captions.}

\item{Fig. 1-}{{\it Bubble Nucleation.}
An open universe emerges through the
bubble nucleation, described
semiclassically by the
Coleman--de Luccia instanton and its continuation
to Lorentzian spacetime.}

\item{Fig. 2-}{{\it Potential for Open Inflation.}}

\refout

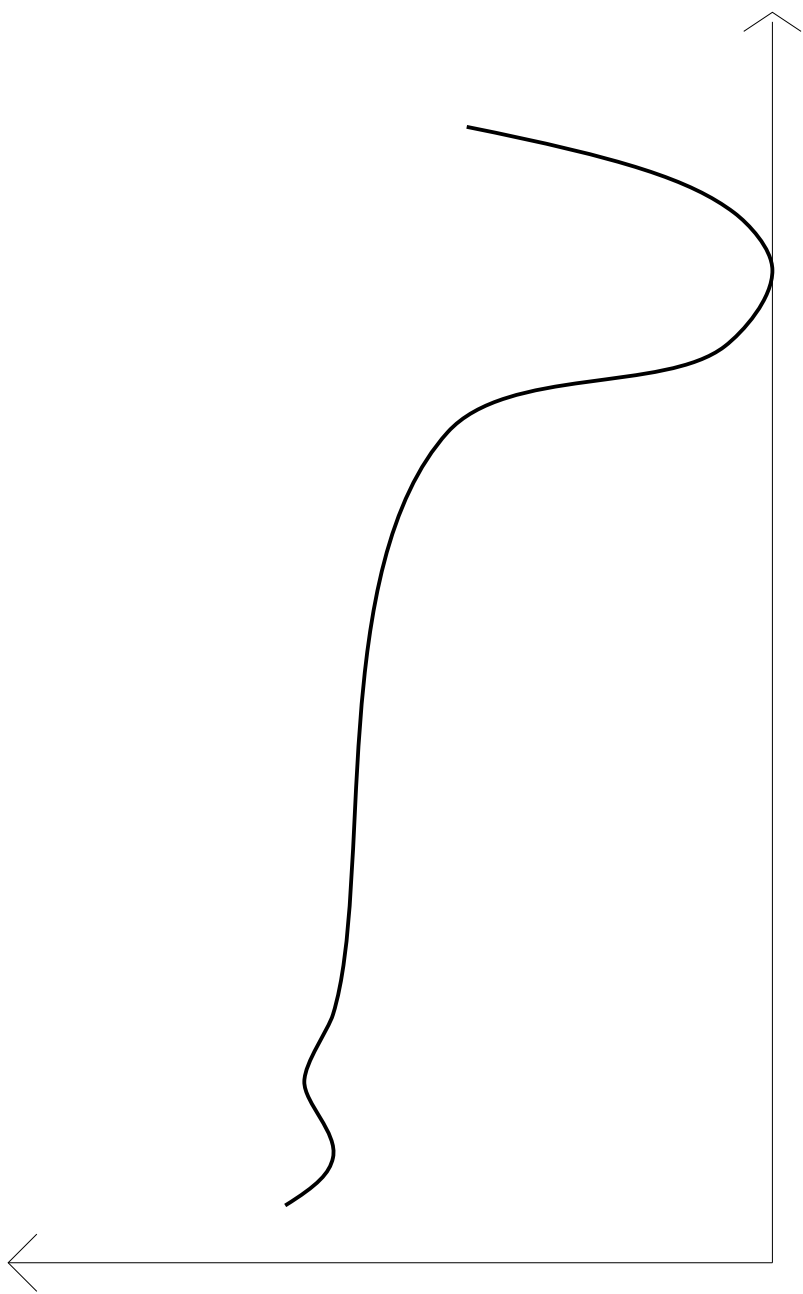
\end